# Interface-Induced Sign Reversal of the Anomalous Hall Effect in Magnetic Topological Insulator Heterostructures

Fei Wang[1, 2, 3, 5], Xuepeng Wang[4,5], Yi-Fan Zhao[1,5], Di Xiao[1], Ling-Jie Zhou[1], Wei Liu[2], Zhidong Zhang[2], Weiwei Zhao[3], Moses H. W. Chan[1], Nitin Samarth[1], Chaoxing Liu[1], Haijun Zhang[4], and Cui-Zu Chang[1]

[1]Department of Physics, The Pennsylvania State University, University Park, PA 16802, USA

[2]Shenyang National Laboratory for Materials Science, Institute of Metal Research, Chinese Academy of Sciences, Shenyang 110016, China

[3]School of Material Science and Engineering, Harbin Institute of Technology, Shenzhen, 518055, China

[4]National Laboratory of Solid-State Microstructures, School of Physics, Nanjing University, Nanjing 210093, China

[5]These authors contributed equally to this work.

Corresponding authors: zhanghj@nju.edu.cn (H. Z.); cxc955@psu.edu (C.-Z. C.).

**The Berry phase picture provides important insights into the electronic properties of condensed matter systems. The intrinsic anomalous Hall (AH) effect can be understood as the consequence of non-zero Berry curvature in momentum space. Here we fabricated TI/magnetic TI heterostructures and found that the sign of the AH effect in the magnetic TI layer can be changed from being positive to negative with increasing the thickness of the top TI layer. Our first-principles calculations show that the built-in electric fields at the TI/magnetic TI interface influence the band structure of the magnetic TI layer, and thus lead to a reconstruction of the Berry curvature in the**

**heterostructure samples. Based on the interface-induced AH effect with a negative sign in TI/V-doped TI bilayer structures, we created an artificial "topological Hall effect"-like feature in the Hall trace of the V-doped TI/TI/Cr-doped TI sandwich heterostructures. Our study provides a new route to create the Berry curvature change in magnetic topological materials that may lead to potential technological applications.**

The Berry phase is a quantum geometrical phase which has provided deep insights into the topological electronic properties of quantum materials [1, 2, 3, 4]. Since the Berry phase encodes the adiabatic evolution of occupied eigen wave functions around the Fermi surface in the first Brillouin zone (BZ) of momentum space [4, 5], it is also important for understanding certain physical phenomena of quantum materials such as the intrinsic anomalous Hall (AH) effect [6]. In the 1950s, Karplus and Luttinger first proposed the idea that an AH effect of intrinsic origin can arise from the properties of the electronic band structure of a material [7]. This connection was reformulated in the language of the Berry phase in the early 2000s [6, 8, 9, 10]. The Berry curvature $\Omega(\vec{k})$ of the occupied Bloch bands is equivalent to an effective magnetic field in momentum space, this can affect the motion of electrons and gives rise to the AH effect in ferromagnetic (FM) materials [6]. Therefore, the Hall conductance ($\sigma_{xy}$) of the intrinsic AH effect in a two-dimensional (2D) FM material can be, in essence calculated from the Fermi-sea integration of the Berry curvature $\Omega$ in its first BZ:

$$\sigma_{xy} = -\frac{e^2}{2\pi h} \int_{BZ} \Omega(\vec{k}) d^2\vec{k} \qquad (1)$$

where $e$ is the elementary charge, $h$ is the Planck constant and $\vec{k}$ is the momentum wavevector. Because the integration of Berry curvature $\Omega$ for occupied Bloch bands in the first BZ of a 2D FM insulator equals to its Chern number $C$ multiple of $2\pi$, i.e. $\int_{BZ} \Omega(\vec{k})d^2\vec{k} = 2\pi C$, $\sigma_{xy}$ is therefore quantized as the following equation:

$$\sigma_{xy} = -\frac{e^2}{h}C \qquad (2)$$

The experimental realization of the quantum anomalous Hall (QAH) effect in Cr-doped (Bi,Sb)$_2$Te$_3$ film, confirming Eq. (2) with $C$ = 1, also establishes convincingly that the intrinsic mechanism is solely responsible for the AH effect in magnetic TIs at least when the chemical potential of the sample is tuned near or located in the magnetic exchange gap [11, 12, 13, 14, 15, 16, 17, 18]. However, when the chemical potential of magnetic TI is tuned away from the magnetic exchange gap and crosses the bulk bands, it is not clear whether the intrinsic AH effect in magnetic TI is still dominant.

The origin of the AH effect in the metallic regime can be studied by exploring the scaling relationship between the AH resistance ($\rho_{yx}$) and the longitudinal resistance ($\rho_{xx}$). A quadratic dependence (i.e. $\rho_{yx} \propto \rho_{xx}^2$) usually indicates that the AH effect is induced by scattering independent mechanisms, i.e. either intrinsic [6] or extrinsic side-jump [19], while a linear dependence (i.e. $\rho_{yx} \propto \rho_{xx}$) can only be a result of extrinsic skew-scattering [20, 21]. It has been theoretically proposed [8] and experimentally demonstrated [22, 23] that the AH effect in metallic diluted magnetic semiconductors (DMSs) (e.g. Mn-doped GaAs) is dominated by intrinsic mechanisms invoking the Berry phase [6]. Since the magnetically doped TI is in principle also a class of DMSs, it is expected that the intrinsic AH contribution is still dominant when the bulk channels are introduced in the magnetic TI. Provided the intrinsic AH contribution is still dominant, it is not clear whether the AH effect and the corresponding Berry curvature $\Omega$ in metallic magnetic TIs can be altered.

In this *Article*, we deposited TI films of different thicknesses on top of the magnetic TI film to form TI/magnetic TI bilayer heterostructures and systematically carry out Hall measurements under different temperatures ($T$) and different gate voltage ($V_g$). We observed a quadratic dependence between $\rho_{yx}$ and $\rho_{xx}$ (i.e. $\rho_{yx} \propto \rho_{xx}^2$) in the metallic regime of the magnetic TI samples, demonstrating the scattering independent origin of the AH effect. By tuning the TI thickness and/or the electric gate voltages, the magnitude and the sign of the AH effect for the magnetic TI heterostructures can be changed. We carried out first principles calculations and attributed this AH effect change to the interface-induced Berry curvature reconstruction, which occurs due to the band structure modulation of the magnetic TI layers induced by built-in electric fields. We then fabricated the magnetic TI/TI/magnetic TI sandwich heterostructures with different dopants and were at first surprised to find a topological Hall (TH) effect-like behavior in the Hall traces. However, our careful analysis of the data shows that this apparent "TH effect" is not a result of the formation of chiral spin textures in the samples but due to the superposition of two AH effects with opposite signs. The interface-induced AH change realized in magnetic TI heterostructures opens a new avenue for systematic studies of the intrinsic AH effect of magnetic topological materials.

**Results**

**Samples structures and intrinsic AH effect in metallic magnetic TI films.** The samples used in this work are $m$ quintuple layers (QL) $Sb_2Te_3$ with $m$ = 0, 1, 2, 3, 4, and 5 on top of 5 QL V-doped $Sb_2Te_3$ bilayers and 5 QL Cr-doped $Sb_2Te_3$/3 QL $Sb_2Te_3$/5 QL V-doped $Sb_2Te_3$ sandwiches. All samples were grown on 0.5 mm thick heat-treated $SrTiO_3$ (111) substrates in molecular beam epitaxy (MBE) chambers. Electrical transport measurements were performed in a Quantum Design Physical Property Measurement System (PPMS) (2 K/9 T) with the magnetic field applied perpendicular to the film plane. The excitation current used in our measurements is 1 μA. Six-terminal Hall bars with a bottom-gate electrode were used for electrical transport studies. The Hall transport results shown here were anti-symmetrized as a function of the magnetic field and the ordinary Hall effect contributions were subtracted.

We first examined the relationship between the intrinsic AH effect and the integration of Berry curvature $\Omega$ of all the occupied Bloch bands in FM materials. As shown in Eq. (1), in the metallic regime of the 2D ferromagnets, the non-quantized Hall conductance $\sigma_{xy}$ can be calculated by integrating the Berry curvature $\Omega$ in the first BZ ($\int \Omega$) but with an opposite sign [3]. When an external magnetic field $B$ is applied perpendicular to the sample, the spontaneously ordered magnetic moments deflects the motion of electrons and the AH effect emerges (**Figs. 1a** and **1b**). When the internal magnetization $M$ points upward, the mobile electrons move to the high potential side for $\int \Omega < 0$, so $\sigma_{xy} > 0$ (**Fig. 1a**) and the intrinsic AH hysteresis exhibits a positive sign (**Fig.1c**). If $\int \Omega > 0$ in the FM materials, the mobile electrons will move to the lower potential side under positive internal magnetization $M$, so $\sigma_{xy} < 0$ (**Fig. 1b**) and the sign of the intrinsic AH effect becomes negative (**Fig. 1d**). Therefore, the intrinsic AH effect can be used as an electrical transport probe to detect the Berry curvature $\Omega$ of the FM materials.

In order to demonstrate the dominance of the intrinsic AH effect in the metallic regime of magnetically doped TI films, we studied the dependence between $\rho_{yx}$ and $\rho_{xx}$ in a series of 5QL V-doped $Sb_2Te_3$ ($Sb_{2-x}V_xTe_3$) samples with different V concentration $x$ ($x \geq$ 0.10). **Figure 1e** shows the $\rho_{yx}$ under zero magnetic field (labeled as $\rho_{yx}(0)$) versus $\rho_{xx}$ under zero magnetic field (labeled as $\rho_{xx}(0)$) curve plotted on a log-log scale in the low-temperature regions (more transport results are shown in Section II of Supplementary Information). This curve is linearly fitted and gives the scaling exponent α in the formula $\rho_{yx}(0) \propto \rho_{xx}^{\alpha}(0)$. We found α ~1.84 ± 0.02, 1.94 ± 0.01, and 1.91 ± 0.02 for $x$ = 0.10, 0.16, and 0.20 samples, respectively. Since the α values are all very close to 2, we conclude the scattering independent (i.e. the intrinsic and/or extrinsic side-jump) mechanisms are primarily responsible for the AH effect in the metallic magnetic TI films [6]. However, since the TI materials have strongly spin-orbit coupled bands [1, 2], the intrinsic contribution is expected to be dominant [6], similar to the AH effect in the well-studied magnetic Heusler and half-Heusler metals [24, 25, 26]. We will next focus on altering this intrinsic AH effect by adding additional undoped TI layer(s) on top of magnetic TI layers.

**Interface-induced AH change in TI/magnetic TI bilayers.** **Figures 2a** to **2f** show the AH hysteresis loops of $m$ QL $Sb_2Te_3$/5 QL $Sb_{1.9}V_{0.1}Te_3$ bilayer samples with $0 \leq m \leq 5$. For the $m$ = 0 and $m$ = 1 samples, the sign of the AH hysteresis loops is positive (i.e. $\rho_{yx} > 0$ for $M > 0$). Under zero magnetic field, $\rho_{yx}(0)$ are ~62 Ω and ~7.5 Ω, respectively (**Figs. 2a** and **2b**). Both values are much lower than the quantized values (~25.8 kΩ), indicating that the chemical potential is crossing the bulk valence bands and the sample is located in the metallic regime [12, 27]. When we increase the thickness ($m \geq 2$) of the $Sb_2Te_3$ layer, the sign of the AH effect reverses and becomes negative (i.e. $\rho_{yx} < 0$ for $M > 0$). The absolute value of $\rho_{yx}(0)$ increases from ~1.2 Ω for the $m$ = 2 sample to ~10 Ω for the $m$ = 4 sample (**Figs. 2c** to **2e**). With further increase in the thickness $m$ of the undoped TI layer, $\rho_{yx}(0)$ decreases due to the introduction of more conduction from the thick undoped TI layer. For the $m$ = 5 sample, the absolute value of $\rho_{yx}(0)$ is ~ 0.5 Ω and the AH hysteresis loop becomes less squared (**Fig. 2f**).

We summarized the temperature dependence of $\rho_{yx}(0)$ for these six samples in **Fig. 2g**. For the $m = 0$ sample, the temperature dependence of $\rho_{yx}(0)$ is monotonic and $\rho_{yx}(0)$ becomes 0 at $T = 40$ K, suggesting the Curie temperature ($T_C$) of 5 QL $Sb_{1.9}V_{0.1}Te_3$ is ~40 K. This temperature dependence is typical in conventional DMS [28] and magnetically doped TI samples [29, 30]. For the $m = 1$ sample, $\rho_{yx}(0)$ first increases and then decreases with temperature showing a maximum at $T =15$ K. $\rho_{yx}(0)$ vanishes at $T = 40$ K. For the $m = 2$ sample, as noted above, $\rho_{yx}(0) < 0$ at $T = 2$ K. With further increase in temperature, the magnitude of $\rho_{yx}(0)$ decreases monotonically and becomes zero at $T = 8$ K. $\rho_{yx}(0)$ of the $m = 3$ sample shows an unusual temperature dependence, specifically it changes its sign from being negative to positive near $T = 11$ K. In other words, this sample shows a linear Hall trace at $T = 11$ K. Note that the linear Hall trace observed here does not mean that the ferromagnetic order is absent because the sample still shows a butterfly magnetoresistance curve (see Section III of Supplementary Information). Instead, $\rho_{yx}(0) = 0$ at $T =11$ K indicates $\int \Omega = 0$ of the occupied Bloch bands in the 3 QL $Sb_2Te_3$/5 QL $Sb_{1.9}V_{0.1}Te_3$ bilayer sample. For the $m = 4$ and $m = 5$ sample, $\rho_{yx}(0) < 0$ in the entire temperature range and its absolute value decreases monotonically with increasing temperature $T$. We note that the butterfly structure of $\rho_{xx}$ in all samples disappears at $T = 40$ K (Fig. S7), suggesting that $T_C$ of all these samples is ~ 40 K and the V atoms in the bottom V-doped TI layer do not diffuse into the top TI layer. We, therefore, concluded the magnitude and the sign changes in these bilayer heterostructures are results of the change in the integration of the Berry curvature $\int \Omega$. This will be discussed in more details below based on the gate tuning of the sample.

**Figures 3a** to **3e** show the magnetic field ($\mu_0H$) dependence of $\rho_{yx}$ under different gate voltage $V_g$s at $T = 11$ K for the $m = 3$ sample. As noted above, $\rho_{yx}$ shows a linear behavior at $V_g = 0$ V (**Fig. 3c**). By applying a negative $V_g$ to introduce hole carriers into the sample, the AH hysteresis with a positive sign appears, suggesting $\int \Omega < 0$ (**Figs. 3a** and **3b**). When applying a positive $V_g$, the electron carriers are introduced, the AH hysteresis with a negative sign appears, suggesting $\int \Omega > 0$ (**Figs. 3d** and **3e**). Therefore, the samples tend to show the AH effect with the positive (negative) sign when injecting the hole (electron) carriers. It is interesting to note that our observation here is opposite to that reported in (Bi,Sb)$_2$Te$_3$/Cr-doped (Bi,Sb)$_2$Te$_3$ bilayer samples [31]. **Figure 3f** shows the 2D color contour plot of $\rho_{yx}(0)$ on the $T$ and $V_g$ plane, which is divided into two regions by a dashed line that extends to zero temperature. Positive $\rho_{yx}(0)$ is favored at higher $T$ and negative $V_g$ with more hole carriers injected, while negative $\rho_{yx}(0)$ is favored at lower $T$ and positive $V_g$ with more electron carriers injected. Therefore, both the magnitude and the sign of $\rho_{yx}$ can also be changed by varying the gating voltage.

**Theoretical analysis of the interface-induced AH change in TI/magnetic TI bilayers.** The magnitude and the sign changes of $\rho_{yx}$ induced by varying the thickness of the top undoped TI layer can be interpreted based on the Berry curvature distribution change in the bottom V-doped Sb$_2$Te$_3$ layer, modulated by the built-in electric fields at the interface between the Sb$_2$Te$_3$ and V-doped Sb$_2$Te$_3$ layers. Prior studies have demonstrated that the V dopants in Sb$_2$Te$_3$ can introduce more hole carriers [12, 32, 33]. In other words, the chemical potential of the V-doped Sb$_2$Te$_3$ layer compared to that of the Sb$_2$Te$_3$ layer is more deeply buried in the bulk valence bands (**Fig. 4a**). When an additional $Sb_2Te_3$ layer is deposited on top of the 5QL V-doped $Sb_2Te_3$ layer to form the $Sb_2Te_3$/V-doped $Sb_2Te_3$ bilayer heterostructures, additional hole carriers in the V-doped $Sb_2Te_3$ layers can move to the $Sb_2Te_3$ layers and thus a built-in electric field is created at the $Sb_2Te_3$/V-doped $Sb_2Te_3$ interface. As a consequence, the chemical potentials of the $Sb_2Te_3$ and V-doped $Sb_2Te_3$ layers are pulled to the same level (**Fig. 4a**). The thicker undoped $Sb_2Te_3$ layer expects to accept more hole carriers from the V-doped $Sb_2Te_3$ layer and thus induce a larger built-in electric field at the $Sb_2Te_3$/V-doped $Sb_2Te_3$ interface.

To simulate the effect induced by the built-in electric field at the TI/magnetic TI interface, we carried out first-principles calculations on the electronic band structures and the corresponding AH conductance $\sigma_{xy}$ of 5 QL $Sb_{1.9}V_{0.1}Te_3$ films with Structure II (i.e. two V atoms located at the third outermost Sb layers, see more details and calculated electronic band structures and the Berry curvature distributions for other three kinds of structures in Section VIII of Supplementary Information) under different electric fields (**Figs. 4b** to **4i**). We see that there is no QAH effect regime for all cases. A trivial gap is formed between the bulk conduction and valence bands, where the Hall conductance $\sigma_{xy}$ is zero. We speculate this gap is likely induced by the hybridization between the top and bottom surface states [34]. We noted that the TI/magnetic TI bilayer structures are not suitable for the realization of the QAH effect because one surface state of the undoped TI layer is always not gapped. By adjusting the magnitude of *E*, we can simulate the observed intrinsic AH effect of the $Sb_2Te_3$/V-doped $Sb_2Te_3$ bilayer heterostructures with different thicknesses of the undoped $Sb_2Te_3$ layers. The calculated AH conductance $\sigma_{xy}$ can be either positive or negative with the chemical potential located at different energy levels (**Figs.** 4**b, 4d, 4f,** and **4h**). The Berry curvature distributions under different $E$ were also calculated with the chemical potential located at different energy levels (i.e. +42 meV, 0 meV, and -30 meV) (**Figs. 4c**, **4e, 4g,** and **4i**). These energy levels are chosen to be within the range of the $V_g$-induced chemical potential shift. Our calculations show that the band crossing and/or anti-crossing (by spin-orbit coupling) in the bulk bands can contribute a large Berry curvature, consistent with prior studies [9, 10]. Once the chemical potential shifts away from the crossing and/or anti-crossing bands or the built-in electric field modulates the crossing and/or anti-crossing bands, the Berry curvature would have a significant change, which can correspondingly change the AH conductance $\sigma_{xy}$ including both magnitude and sign.

For the 5 QL V-doped $Sb_2Te_3$ films without the electric field (i.e. $E$ =0 V/Å), the spatial inversion is preserved, and the Berry curvature distribution shows an exact hexagonal pattern (**Fig. 4c**). We found that there is a narrow energy region with a positive AH conductance $\sigma_{xy}$ near the Fermi level which originates from the large negative Berry curvature induced by the appearance of the band anti-crossing in the calculated electronic band structures (**Fig. 4b**) [9, 10]. Once electric fields are introduced, the spatial inversion symmetry is broken and the Berry curvature distribution consequently loses the hexagonal pattern (**Figs. 4e, 4g,** and **4i**). We note that the electric field $E$ can evidently tune the position of the band anti-crossing. For example, within the energy region of the positive AH conductance $\sigma_{xy}$, the gap of band anti-crossing along Γ−K' first close and then reopen under the electric fields, which changes the Berry curvature from negative to positive. Larger electric fields (e.g. $E$ =0.001 V/Å and 0.0015 V/Å) tunes $\sigma_{xy}$ from being positive to negative (**Figs. 4f** to **4i**). Therefore, our first-principles calculations demonstrate that the built-in electric fields play a key role in the reconstruction of the Berry curvature and the magnitude and sign change of $\sigma_{xy}$ in $Sb_2Te_3$/V-doped $Sb_2Te_3$ bilayer samples.

Next, we discuss whether the chemical potential of $Sb_2Te_3$/V-doped $Sb_2Te_3$ bilayer samples in experiments could locate into or near the AH sign reversal region of our first-principles calculations. We made the following rough estimation. From our Hall results, the carrier densities of the $m = 0$ to $m = 5$ samples at $V_g$ =0V (**Fig. 2**) are 1.04, 3.54, 5.93, 1.67, 2.46, and $1.74 \times 10^{13}$ cm$^{-2}$, respectively. By combining with the calculated density of states of the 5 QL $Sb_{1.9}V_{0.1}Te_3$ films under $E = 0$ V/Å, we estimated the positions of the chemical potential are ~0.07 eV, ~ 0.13 eV, ~0.19 eV, ~0.09 eV, ~0.11 eV, and 0.09 eV below the top of bulk valence bands. We found that the estimated chemical potentials are slightly higher than the AH sign reversal regime induced by the built-in electric fields except for the $m$ =2 sample. We attributed this small difference to a reasonable error bar of the estimation. There are two kinds of approximations made in our first-principles calculations. The first approximation is the assumption that the carriers uniformly distributed in the undoped $Sb_2Te_3$/V-doped $Sb_2Te_3$ heterostructure samples, while the carrier concentrations in real bilayer samples are different in undoped $Sb_2Te_3$ and V-doped $Sb_2Te_3$ layers (**Fig. 4a**), as discussed above. The second approximation is to use the supercell model to simulate the V doping in our calculations, in which V doping atoms are assumed to distribute periodically, thus forming bands near the Fermi energy. A significant portion of bulk valence bands comes from the contribution of V atom *d* orbitals near the Fermi energy while V atoms in the real V-doped TI samples should be distributed completely randomly and the atomic orbitals from V atoms cannot form bands.

This means that some hole carriers that occupied the V atom orbitals should be localized and cannot contribute to the transport measurements. Therefore, it is acceptable to consider that the chemical potential in real samples locates within or near the AH sign reversal region in our theoretical calculations. In our experiments, the $m = 0$ sample without the built-in electric field always shows a positive AH sign, which further supports our interpretation.

As noted above, the electron from the undoped TI to magnetic TI layers yields a built-in electric field at the interface. This built-in electric field moves the chemical potential of the magnetic TI layer upward and at the same time reconstructs the electronic band structure, as well as the corresponding Berry curvature distribution. We therefore conclude that the interface-induced AH effect change in TI/magnetic TI bilayer sample is indeed from the Berry curvature distribution change. For the $m = 4$ sample, the AH resistance reaches a maximum value (**Fig. 2**), so we roughly estimate the effective penetration length of the built-in electric field as 4 QL (i.e. 4 nm) in our TI/magnetic TI bilayer samples. In other words, any undoped TI layer greater than 4 QL will not further contribute to the formation of the built-in electric field, but the better conduction of the thick undoped TI layer will shunt the current going through the bottom magnetic TI layer and thus the AH effect will decrease and finally disappear. Note that some valence bands induced by random V dopants are localized and does not contribute the AH conductance $\sigma_{xy}$, which should be overestimated in our calculations. Moreover, the carriers in the real samples are inevitably scattered by the impurities and defects and thus have a finite relaxation time [6], which would reduce the AH velocity and significantly suppress the intrinsic AH conductance $\sigma_{xy}$ in the dirty limit [35]. Therefore, the calculated values of $\sigma_{xy}$ here are generally larger than the experimental values shown in **Fig. 2**.

The temperature dependence of the AH resistance $\rho_{yx}$ (**Fig. 2f**) can also be qualitatively interpreted by the above picture. For the $m =1, 2$, and $3$ samples, a large number of holes can be thermally excited with increasing temperature in both V-doped and undoped $Sb_2Te_3$ layers, thus weakening the electron transfer effect between these two layers, so $\rho_{yx}$ of the $Sb_2Te_3$/V-doped $Sb_2Te_3$ heterostructure will become more similar to the $m = 0$ sample. The weakening of charge transfer effect explains the sign change of the AH resistance $\rho_{yx}$ for the $m = 3$ sample with increasing temperatures. When the temperature approaches towards $T_C$, the magnitude of $\rho_{yx}$ always decreases towards zero due to the disappearance of magnetization.

In the following, we discuss the different roles of the growth of the undoped TI layer and the application of the bottom gate voltage for the AH change in our experiments. By varying the thickness of the undoped TI layer, it mainly affects the magnitude of the built-in electric fields and thus influences the electronic band structures and Berry curvature distribution, while the application of the gate voltage is primarily to induce the chemical potential shift. In addition, the metallic properties of highly hole-doped $Sb_2Te_3$ or V-doped $Sb_2Te_3$ samples might affect the distribution of the electric fields in TI/magnetic TI bilayer samples, i.e. the built-in electric field in the $m$ QL $Sb_2Te_3$/5QL V-doped $Sb_2Te_3$ bilayers is very likely located only at the interface regime. Based on the above estimated chemical potential positions of different bilayer samples (Fig. 2) and chemical potential shift of the 3QL $Sb_2Te_3$/5QL V-doped $Sb_2Te_3$ sample (Fig.3). We found both values are consistent with our built-in electric field picture, so our theoretical analysis qualitatively explains our experimental results.

**Artificial "TH effect" in Cr-doped TI/TI/V-doped TI trilayer samples.** Since the Cr-doped TI layer and Cr-doped TI/TI bilayer usually show the AH effect with a positive sign [30] (see Section VI of Supplementary Information), the realization of the AH effect with a negative sign in TI/V-doped TI bilayer structures provides us an opportunity to study two decoupled ferromagnetic orders with opposite AH signs and different coercive fields ($\mu_0 H_c$). Therefore, we fabricated the 5 QL Cr-doped $Sb_2Te_3$/3 QL $Sb_2Te_3$/5 QL V-doped $Sb_2Te_3$ sandwich heterostructures (**Fig. 5**) and systematically studied its Hall traces at different temperatures. In our sandwich heterostructures, the middle 3QL $Sb_2Te_3$ plays the following duo roles: (*i*) it weakens the interlayer coupling effects between the top and bottom magnetic TI layers; (*ii*) it drives the bottom V-doped TI shows the AH effect with negative sign, as discussed above. In Hall measurements, the two AH effects with opposite AH signs and different $\mu_0 H_c$ can be added together to form unusual AH hysteresis loops. When the absolute value of $\rho_{yx}$ in Cr-doped TI is larger than that of V-doped TI, the AH effect of the Cr-doped TI layer is dominant, and the total AH hysteresis loop shows a positive sign (**Fig. 5a**), we named this as "Type 1" AH hysteresis loop. When the absolute value of $\rho_{yx}$ in Cr-doped TI is less than that of V-doped TI, the AH effect of the V-doped TI layer is dominant, and the total AH hysteresis loop shows a negative sign (**Fig. 5b**), we named it as "Type 2" AH loop. In both cases, two anti-symmetric humps appear when the external magnetic field $\mu_0 H$ is between $\mu_0 H_{c1}$ and $\mu_0 H_{c2}$. Here $\mu_0 H_{c1}$ and $\mu_0 H_{c2}$ are the coercive fields of the top Cr-doped TI and the bottom V-doped TI layers, respectively. A "hump" feature in the AH hysteresis loop can result from the formation of chiral spin textures in samples and is usually known as the TH effect [36, 37, 38, 39]. The hump feature observed in our sandwich samples, on the other hand, is the result of the superposition of two AH effect with opposite signs.

We performed Hall measurements on the 3QL $Sb_2Te_3$/5QL $Sb_{1.92}V_{0.08}Te_3$ and 5QL $Sb_{1.84}Cr_{0.16}Te_3$/3QL $Sb_2Te_3$ bilayer samples. As discussed above, the former sample does show the AH effect with a negative sign (**Figs. 5d** and **5g**), while the latter shows the AH effect with the positive sign (**Figs. 5e** and **5h**), consistent with prior studies [29, 30] and our theoretical calculations (see Section IX of Supplementary Information). Because the temperature dependences of $\rho_{yx}$ in these two bilayer samples are different, both types of AH loops can be realized in the 5QL $Sb_{1.84}Cr_{0.16}Te_3$/3QL $Sb_2Te_3$/5QL $Sb_{1.92}V_{0.08}Te_3$ samples in different temperature regions (see Section VI of Supplementary Information). Since the absolute value of $\rho_{yx}$ in 5QL $Sb_{1.84}Cr_{0.16}Te_3$/3QL $Sb_2Te_3$ is larger than that of 3QL $Sb_2Te_3$/5QL $Sb_{1.92}V_{0.08}Te_3$ at $T$ =2 K, "Type 1" AH loop is observed (**Fig. 5f**). At $T$ =12 K, the absolute value of $\rho_{yx}$ in 5QL $Sb_{1.84}Cr_{0.16}Te_3$/3QL $Sb_2Te_3$ is comparable to or less than that of 3QL $Sb_2Te_3$/5QL $Sb_{1.92}V_{0.08}Te_3$, therefore "Type 2" AH loop appears (**Fig. 5i**). All observations here are consistent with our analysis in **Figs. 5a** and **5b**, confirming that the TH effect-like hump feature observed in our Cr-doped TI/TI/V-doped TI sandwiches are indeed from the superposition of two AH effects with opposite signs. Our findings here, together with two prior studies [40, 41] suggest humps/dips features along with the AH effect observed in a number of bilayer heterostructure samples might not be induced by the formation of the chiral spin textures in samples [31, 42, 43, 44, 45]. The TH-like hump feature observed in these bilayer heterostructures is likely also due to the coexistence of two decoupled ferromagnetic orders. These two ferromagnetic orders could be different ferromagnetisms in bulk and surface layers [31, 45] or different ferromagnetisms on two surfaces of the magnetic layer (i.e. $SrRuO_3$) [42, 43, 44].

To summarize, we observed the sign reversal of the AH effect in magnetic TI samples by depositing an undoped TI layer on top to form bilayer heterostructures. Our observation is well interpreted by our first-principles calculations based on the model of charge transfer through the interface between TI and magnetic TI layers. We fabricated the magnetic TI/TI/magnetic TI sandwich sample with different dopants on two surfaces and observed two atypical humps in the AH loops. We demonstrated that the unusual AH hysteresis loops are induced by the superposition of two AH effects with opposite signs rather than the formation of the chiral spin textures in these samples. Our work provides a new route for the "engineering" of the AH effect in magnetic topological materials and insights into the underlying mechanism for the TH effect-like humps observed in nonhomogeneous materials.

**Methods**

**MBE growth of magnetic TI heterostructures.**

The magnetic TI films and heterostructures were grown on heat-treated 0.5 mm thick insulating $SrTiO_3$(111) substrates using two MBE chambers (Omicron Lab 10 and Vecco Applied EPI 620) with a base pressure of $2.0 \times 10^{-10}$ mbar. Prior to sample growth, the $SrTiO_3$ substrates are first degassed at ~600°C for 1h. High-purity Sb(6N), Te(6N), V(5N), and Cr(5N) are evaporated from Knudsen cells. During the growth of the magnetic TI films, the substrate is maintained at ~240 °C. The flux ratio of Te/(Sb+V/Cr) is set to be >10 to avoid possible Te deficiency in samples. The growth rate of magnetic TI or TI films is around ~0.2 QL/min. Finally, the samples are capped with a 10 nm Te layer to prevent their degradation during the *ex-situ* electrical transport measurements. The V concentration in V-doped $Sb_2Te_3$ films is tuned by adjusting the temperature of the V effusion cell. The $x = 0.08$ to $0.2$ samples correspond to the V cell temperatures from 1440 °C to 1490 °C. For the $Sb_{1.84}Cr_{0.16}Te_3$ sample, the temperature of the Cr effusion cell is 1160 °C. The Cr and V concentration values in this work are calibrated by the plot in our prior work [12].

**Electrical transport measurements.**

All samples for transport measurements were scratched into Hall bar geometries (~ 1 mm × 0.5 mm) [46]. The bottom gate electrode is made by pressing an indium foil on the backside of the $SrTiO_3$ (111) substrate. Note that for all Hall data shown in this paper, unless pointed out, the ordinary Hall term has been subtracted by fitting the linear slope of the Hall data in a high magnetic field regime.

**First-principles calculations.**

The V atom is doped in $2\times2\times5$ supercell of $Sb_2Te_3$ (5 QL thickness) without breaking the spatial reversal symmetry. The concentration of vanadium is about 5% (i.e. $Sb_{1.9}V_{0.1}Te_3$). The lattice parameters and the positions of atoms were fixed as the bulk ones, and we relaxed the dopants and atoms in the nearest neighbor of the dopants with an energy tolerance of $10^{-5}$ eV. A 20 Å vacuum layer is placed above the sample. The density functional theory (DFT) calculations are performed via the Vienna Ab-initio Simulation Package (VASP) with exchange-correlation energy described by the generalized gradient approximation (GGA) in the Perdew-Burke-Ernzerhof (PBE) parametrization. We set the energy cutoff to ~300 eV for the plane-wave expansion. The van der Waals (vdW) correction is also included with the optB86b-vdW type in the relaxation process. An $8\times8\times1$ Γ-centered *k*-mesh is used in structural relaxation with the energy tolerance converged to $10^{-6}$ eV. To consider strongly correlated $3d$ electrons of the V atoms, an LDA+U correction is implemented in the whole calculation with the parameter $U$ = 2.7 eV and $J$ = 0.7 eV. The spin-orbit coupling effect is included in band structure calculations. The AH conductance $\sigma_{xy}$ is calculated in Wannier representation via the Wannier90 code. In the projection from Bloch representation to Wannier representation, considering the computation capability, the Γ-centered $k$-mesh is set to 5×5×1. To include the disorder effect in the magnetic TI samples, a 0.03eV energy range is averaged in the calculated Hall conductance $\sigma_{xy}$. The maximal energy range is estimated by $E_{range} = \frac{\hbar}{2\tau} = 0.09$eV, where $\tau$ is the relaxation time estimated by the carrier mobility data from experiments and the effective mass of valence band edge from the $m$ = 0 DFT calculations.

**Acknowledgments**

The authors would like to thank W. D. Wu, D. Xiao, and X. D. Xu for helpful discussions. C. Z. C. acknowledges support from ARO Young Investigator Program Award (W911NF1810198) and the Gordon and Betty Moore Foundation's EPiQS Initiative (Grant GBMF9063 to C.Z.C.). H. Z. acknowledges support from the National Natural Science Foundation of China (11674165 and 11834006), the Fok Ying-Tung Education Foundation of China (161006), and the Fundamental Research Funds for the Central Universities (020414380149). N. S. acknowledges support from Penn State 2DCC-MIP under NSF grant DMR-1539916. Support for transport measurements and data analysis is provided by the DOE grant (DE-SC0019064). F. W., W. L. and Z. D. Z. acknowledges support from the National Key R & D Program of China (2017YFA0206302). C.X.L. also acknowledges support from the ONR grant (No.

N00014-18-1-2793) and Kaufman New Initiative research grant KA2018-98553 of the Pittsburgh Foundation.

**Author contributions**

C. Z. C. conceived and designed the experiment. Y. F. Z. and D. X. grew the magnetic TI films and heterostructure samples with the help of N. S. and C. Z. C.. F. W., Y. F. Z. and L. Z. carried out the PPMS transport measurement with the help of M. H. W. C. and C. Z. C.. X. W, C. L., and H. Z. provided the theoretical support and did the first-principles calculations. F. W., X. W., C. L., H. Z., and C. Z. C. analyzed the data and wrote the manuscript with contributions from all authors.

**Additional information**

Supplementary information is available in the online version of the paper. Reprints and permissions information is available online at www.nature.com/reprints.

Correspondence and requests for materials should be addressed to H. Z. or C. Z. C.

**Competing interests**

The authors declare no competing interests.

**Data availability**

The data that support the findings of this study are available from H. Zhang or C. -Z. Chang upon reasonable request.

**Figures and figure captions:**

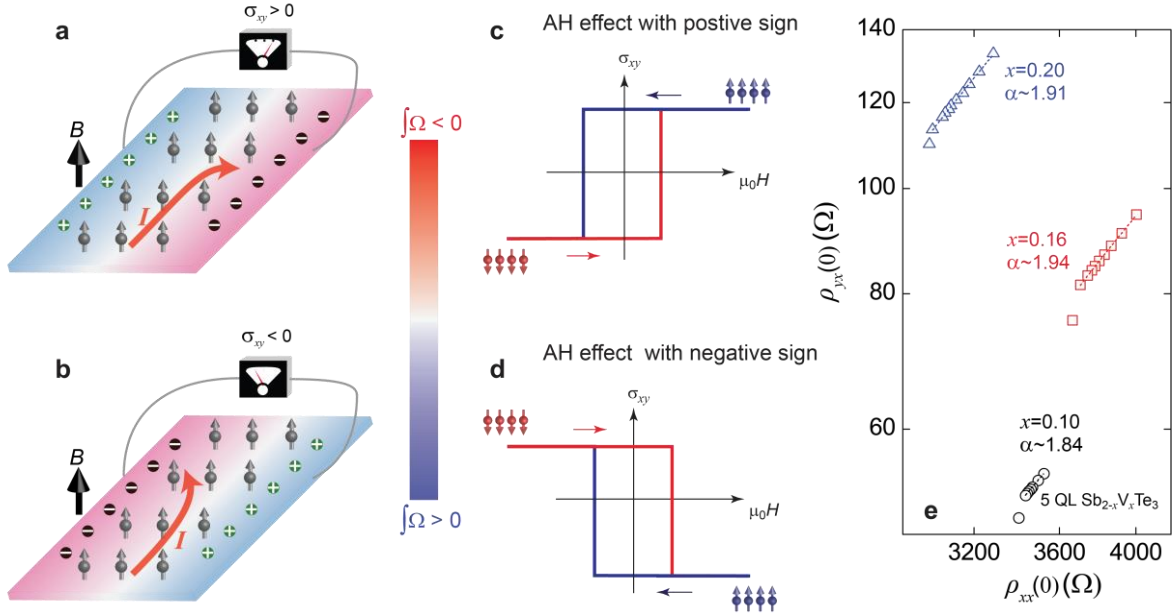

**Figure 1 | Intrinsic anomalous Hall (AH) effect and Berry curvature Ω.** (a,b) The intrinsic AH effect formed in the ferromagnetic materials with positive (a) and negative (b) Berry curvature Ω integration of the occupied Bloch bands when the internal magnetization points upward. *I* is the applied excitation current. ⊕ and ⊖ are hole and electron carriers, respectively. Bule and red colors indicate high and low electrical potentials. (c, d) The corresponding AH hysteresis in the ferromagnetic materials with positive (c) and negative (d) Berry curvature Ω integration of the occupied bands. The sign and value of the AH effect are determined by the sign and magnitude of Berry curvature Ω integration of occupied bands. The arrows indicate the magnetic field sweep directions. The color bar is for (a-d). (e) A plot of the AH resistance at zero magnetic field $\rho_{yx}(0)$ as a function of the longitudinal resistance at zero magnetic field $\rho_{xx}(0)$ on a log-log scale for three 5 QL $Sb_{2-x}V_xTe_3$ samples ($x$=0.10, 0.16, and 0.20). The dashed lines are the fits using $\rho_{yx} \propto \rho_{xx}^{\alpha}$, $\alpha$ =1.84 ±0.02, 1.94±0.01, and 1.91±0.02 for $x$ = 0.10, 0.16, and 0.20 samples, respectively.

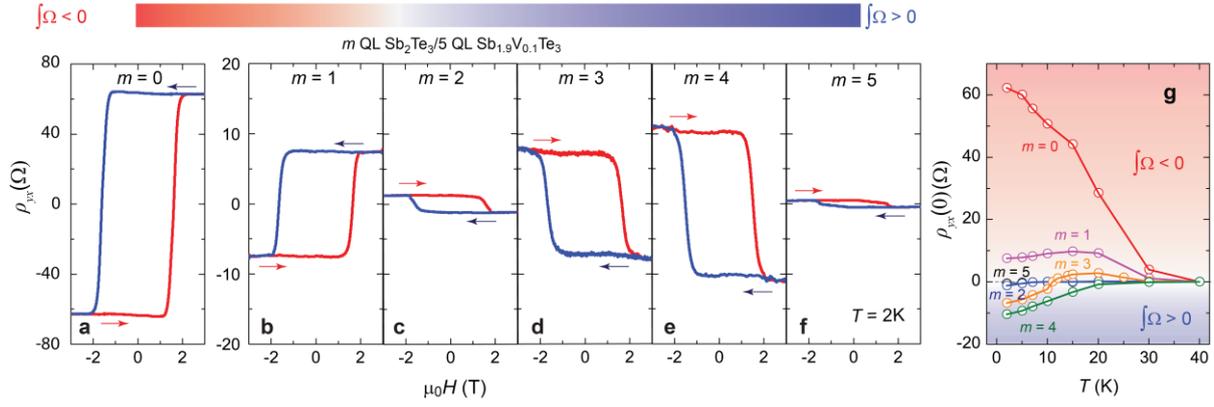

**Figure 2 | Interface-induced sign reversal of the AH effect in TI/magnetic TI heterostructures.** (a-e) Magnetic field ($\mu_0 H$) dependence of the Hall resistance $\rho_{yx}$ in $m$ QL $Sb_2Te_3$/5 QL $Sb_{1.9}V_{0.1}Te_3$ bilayer heterostructures. $m = 0$ (a); $m = 1$ (b); $m = 2$ (c); $m = 3$ (d); $m = 4$ (e); $m = 5$ (f). The arrows indicate the magnetic field sweep directions. All measurements were taken at $T = 2$ K. The linear contribution from the ordinary Hall effect was subtracted. (g) Temperature dependence of the Hall resistance at zero magnetic field $\rho_{yx}(0)$ in $m$ QL $Sb_2Te_3$/5 QL $Sb_{1.9}V_{0.1}Te_3$ bilayer heterostructures.

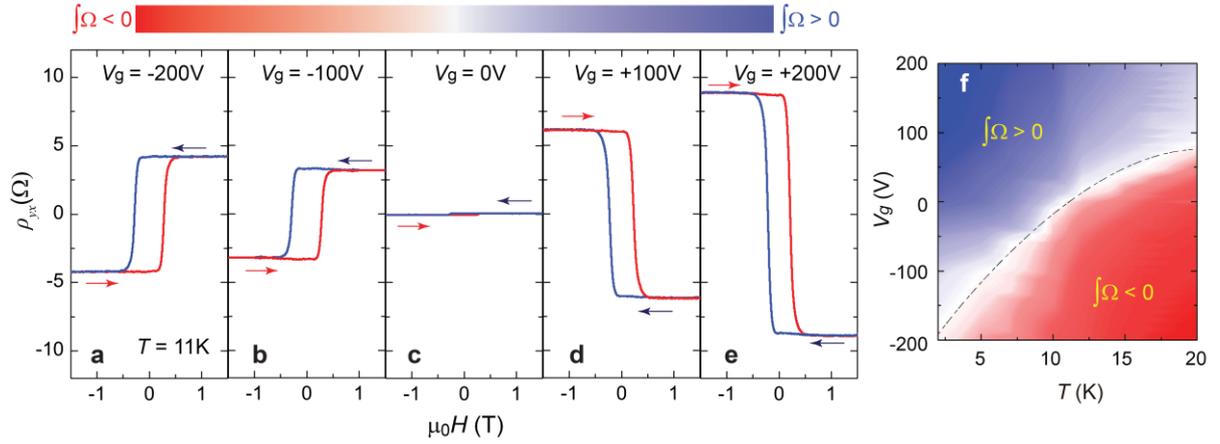

**Figure 3 | Electric field control of the AH change in TI/magnetic TI bilayer heterostructures.** (a-e) $\mu_0 H$ dependence of $\rho_{yx}$ in 3 QL $Sb_2Te_3$/5 QL $Sb_{1.9}V_{0.1}Te_3$ bilayer heterostructure at different gates $V_g$. $V_g = -200V$ (a); $V_g = -100V$ (b); $V_g = 0V$ (c); $V_g = +100V$ (d); $V_g = +200V$ (e). The arrows indicate the magnetic field sweep directions. All measurements were taken at $T = 11$ K. The linear contribution from the ordinary Hall effect was subtracted. (f) 2D color contour plot of $\rho_{yx}$ (0) as a function of both $T$ and $V_g$ in 3 QL $Sb_2Te_3$/5 QL $Sb_{1.9}V_{0.1}Te_3$ bilayer heterostructure.

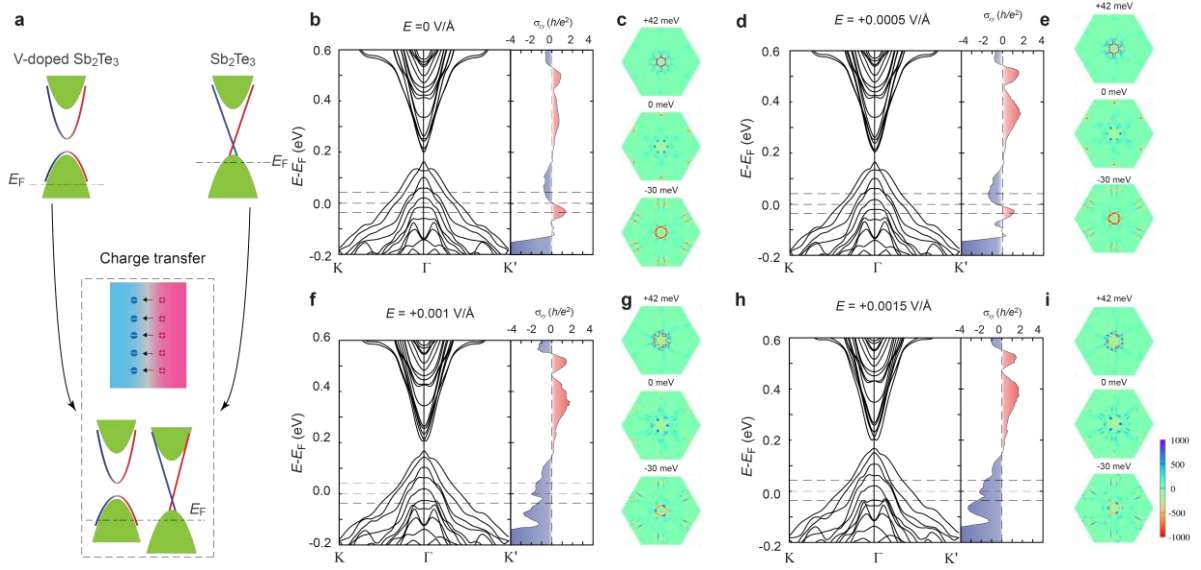

**Figure 4| Schematic of the charge transfer between the TI and magnetic TI layers and calculated band structures and Berry curvature distribution of the magnetic TI films under various electric fields.** (a) Schematic of the charge transfer through the interface between the $Sb_2Te_3$ and V-doped $Sb_2Te_3$ layers. The V-doped $Sb_2Te_3$ layer has more hole carries than the $Sb_2Te_3$ layer. A built-in electric field is generated at the $Sb_2Te_3$/V-doped $Sb_2Te_3$ interface. (b, d, f, h) The calculated band structures (left) and corresponding Hall conductance $\sigma_{xy}$ (right) of 5 QL $Sb_{1.9}V_{0.1}Te_3$ films under electric field $E = 0$ V/Å (b), $E = 0.0005$ V/Å (d), $E = 0.001$ V/Å (f), and $E = 0.0015$ V/Å (h). (c, e, g, i) The calculated Berry curvature distribution of 5 QL $Sb_{1.9}V_{0.1}Te_3$ under electric field $E = 0$ V/Å (c), $E = 0.0005$ V/Å (e), $E = 0.001$ V/Å (g), and $E = 0.0015$ V/Å (i). The chemical potential positions are slightly shifted (-30 meV to 0 meV to 42 meV) to simulate the influence induced by the charge transfer and/or the gating effect in experiments.

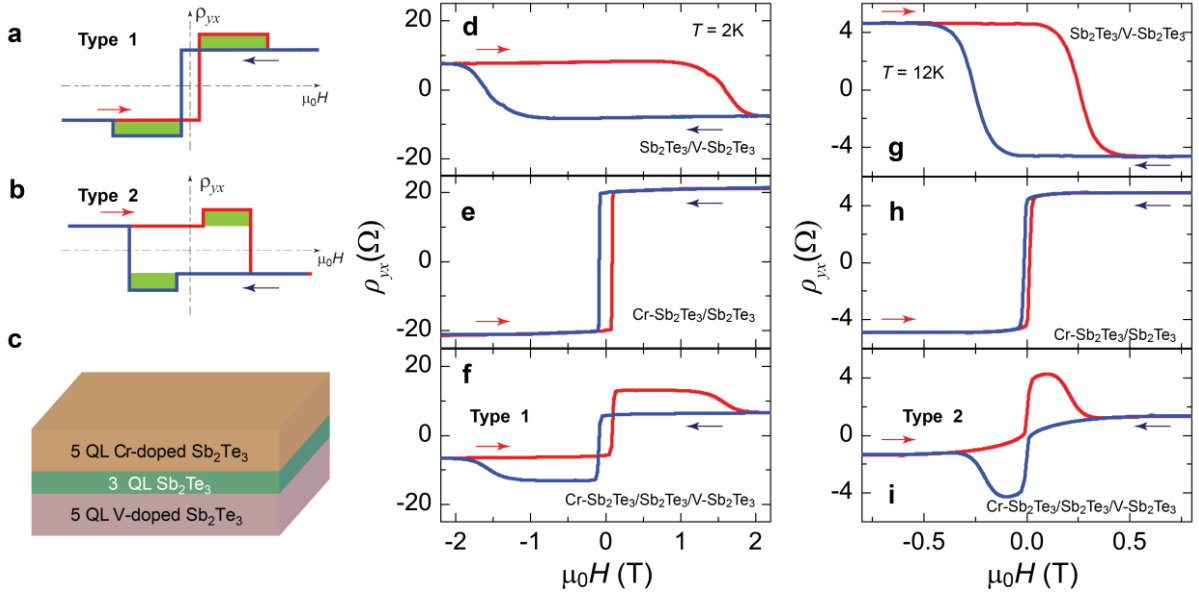

**Figure 5 | The artificial "TH effect" in magnetic TI/TI/magnetic TI sandwich heterostructures.** (a, b) Two types of artificial "TH effect" formed by the parallel connection of two AH effects with opposite signs. Type 1 is constructed by the AH effect with the larger coercive field ($\mu_0 H_c$) showing smaller $\rho_{yx}$, while Type 2 is constructed by the AH effect with the larger $\mu_0 H_c$ showing larger $\rho_{yx}$. (c) Schematic for the 5 QL Cr-doped $Sb_2Te_3$/3 QL $Sb_2Te_3$/5 QL V-doped $Sb_2Te_3$ sandwich sample. (d, g) $\mu_0 H$ dependence of $\rho_{yx}$ of the 3 QL $Sb_2Te_3$/5QL $Sb_{1.92}V_{0.08}Te_3$ bilayer heterostructure measured at $T$ = 2 K (d) and $T$ = 12 K (g). (e, h) $\mu_0 H$ dependence of $\rho_{yx}$ of the 5QL $Sb_{1.84}Cr_{0.16}Te_3$/3 QL $Sb_2Te_3$ bilayer heterostructure measured at $T$ = 2 K (e) and $T$ = 12 K (h). (f, i) $\mu_0 H$ dependence of $\rho_{yx}$ of the 5QL $Sb_{1.84}Cr_{0.16}Te_3$/3 QL $Sb_2Te_3$/ 5QL $Sb_{1.92}V_{0.08}Te_3$ sandwich heterostructure measured at $T$ = 2 K (f) and $T$ = 12 K (i). The arrows indicate the magnetic field sweep directions.